\documentclass[twocolumn,showpacs,preprintnumbers,amsmah,amssymb]{revtex4}
\usepackage{graphicx}
\usepackage{dcolumn}
\usepackage{bm}

\begin{document} 

\title{Dual Behavior of Antiferromagnetic  Uncompensated Spins in NiFe/IrMn Exchange Biased Bilayers}

\author{ S. K. Mishra, F. Radu$^{\ast}$, S. Valencia, D. Schmitz, E. Schierle,  H. A. D\"urr,  and W. Eberhardt}
\affiliation{ Helmholtz-Zentrum f\"ur Materialien und Energie,
Albert-Einstein-Str. 15, D-12489, Berlin, Germany}
\date{\today}

\begin{abstract}
We present a comprehensive study of  the  exchange bias effect in
a model system. Through numerical analysis of the exchange bias
and coercive fields as a function of the antiferromagnetic layer
thickness we deduce the absolute value of the averaged anisotropy
constant of the antiferromagnet. We show that the anisotropy of
IrMn exhibits a finite size effect as a function of thickness. The
interfacial spin disorder involved in the data analysis is further
supported by the observation of the dual behavior of the
interfacial uncompensated spins. Utilizing soft x-ray resonant
magnetic  reflectometry  we have observed that the
antiferromagnetic uncompensated spins are dominantly frozen with
nearly no rotating spins due to the chemical intermixing, which
correlates to the inferred mechanism for the exchange bias.

\end{abstract}
 \maketitle


The tremendous advances of spintronics research   initiated by
 the discovery of interlayer exchange
coupling~\cite{IEC-Grunberg-1988} and giant magnetoresistance
\cite{GMR-Baibich-1988,GMR-Binasch-1989}  uses extensively the
exchange bias (EB) effect to control the magnetization of
ferromagnetic components. This is a consequence of the direct
exchange at the interface between  feromagnetic and
antifereomagnetic layers and/or nanoscale heterostructures  which
causes a shift and a broadening of the hysteresis loop of the
ferromagnet. This effect which was engineered by nature a few
billion years ago~\cite{mcenroe:2007}, was experimentally
discovered 60 years ago by Meiklejohn and Bean (M\&B)
~\cite{bean:1956} when studying Co particles embedded in their
natural oxide (CoO) matrix. Extensive experimental and theoretical
studies of the EB effect provide now sufficient understanding for
utilizing it as a probe for further fundamental
research~\cite{iskhakov:2004,salabas:2006,lenz:2007,chen:2009}.


The EB and coercive fields of the biased ferromagnet (F) are
determined essentially by the magnetic properties of the adjacent
antiferromagnet (AF) and interfacial spin structure. Initially,
the antiferromagnet was considered to be ideally rigid under the
torque exerted during the reversal of the ferromagnetic
layer~\cite{bean:1956}. Soon afterwards, this constraint has been
lifted allowing the AF spins to rotate as a whole during the
magnetization reversal of the ferromagnet~\cite{meiklejohn:1962}.
The bulk AF spins may be displaced from their rigid orientation or
they even can reverse under the torque exerted by the interfacial
coupling. This leads to an onset temperature (blocking
temperature) and AF critical thickness for the EB to occur. These
parameters  are determined by the anisotropy constant of the AF
layer as well as by the nature of the interfacial coupling.

Most recently, yet another proximity effect is being
experimentally unveiled for the interface of EB
bilayers~\cite{ohldag:2003,roy:2005,radu:jmmm:2006,bruck:2008,blackburn:2008,radu:2009,mishra:2009}:
the proximity of the F layer leads to depth uncompensated (UC)
interfacial AF spins which may be loose and frozen. They affect
the interfacial coupling and mediate coercivity in the F layer,
therefore, contributing essentially to the understanding of the EB
effect.

In this Letter we  explore the dependence of the EB coercive
fields on the anisotropy of the AF layer. Through numerical
analysis of the phase diagram we determine the variation of the
AF anisotropy constant as a function of the nanoscale AF
thickness, which exhibits a finite size effect. We show
that this  is a robust and unique capability of EB effect.
We further demonstrate with soft-x-ray resonant magnetic
reflectometry (XRMR) that for sputter-deposited NiFe/IrMn bilayers
an insignificant chemical intermixing minimizes the amount of UC
AF spins rotating with the F. We show that at nearly ideal
exchange biased interfaces, the amount of frozen UC AF spins
dominates and displays a characteristic depth dependence near the
interface. Almost ideal interfaces are also the basis for
extracting the influence of the AF magnetic anisotropy on the
development of EB.

To further provide the confidence in the underlying EB mechanism,
we provide a consistent correspondence between the dual behavior
of UC AF spins components studied by XRMR and interfacial
parameters contributing to the numerical analysis of EB.

A series of specimens Si (100)/SiO$_{2}$/Cu (50~\AA)/
Ni$_{81}$Fe$_{19}$ (75~\AA)/ Ir$_{20}$Mn$_{80}$ (t$_{AF}$ =0, 10,
15, 20, 25 35~\AA)/Cu (25~\AA) were grown on thermally oxidized Si
wafers by using the dc magnetron sputtering technique. The base
pressure in the sputtering chamber was better than
2$\times$10$^{-8}$ mbar. The partial Ar pressure during growth was
set to a minimum value of 1.5$\times$10$^{-3}$ mbar. During
growth, the substrates were intentionally kept at room temperature
(RT) in order to avoid any additional thermal interdiffusion at
the F/AF interface. The uniaxial magnetic anisotropy was induced
in the F layer by applying an in-situ external magnetic field of
2~kOe parallel oriented with respect to the film surface. This
saturated ferromagnetic state provides the means to further induce
the unidirectional anisotropy into the AF layer during growth. As
a seed and capping layer we
 used a {50~\AA}  and {25~\AA}  thick Cu layers, respectively. The
excellent match between the lattice parameter of Cu and
Ni$_{81}$Fe$_{19}$ promotes a low interfacial roughness which is
required for high quality EB bilayers. The thicknesses of the
samples were quartz calibrated and verified with x-ray
reflectivity.

\begin{figure}[!ht]

          \includegraphics[clip=true,keepaspectratio=true,width=1\linewidth]{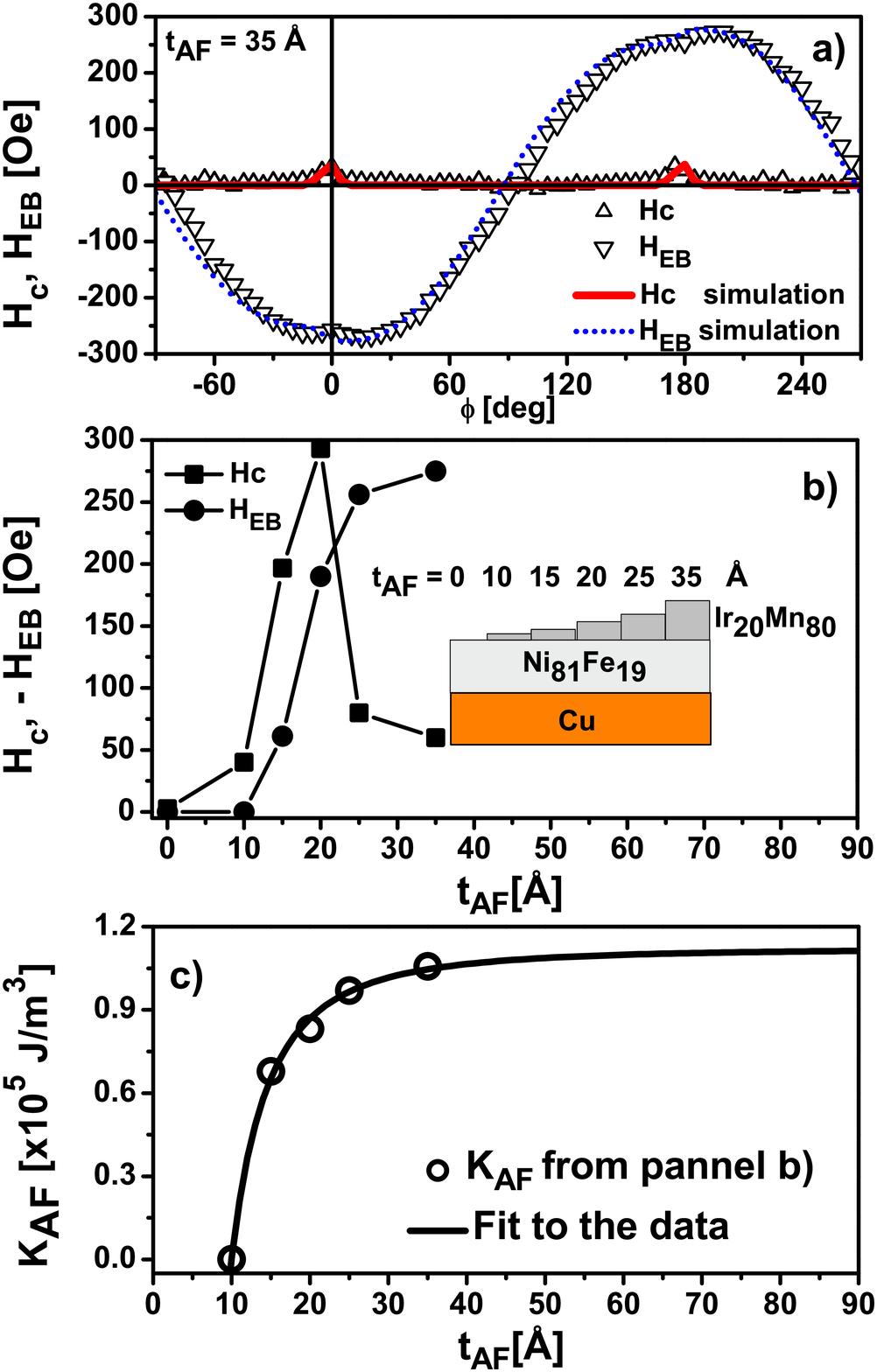}
          \caption{Panel a). The dependence of the
          H$_{EB}$ and the  H$_{c}$ as a
          function azimuthal angle $\phi$ for a representative sample with
          t$_{AF}$=35~\AA. Solid line and dotted line are numerical simulations (see
          text). Panel b). The thickness dependence of the
          H$_{EB}$ and H$_{c}$ (phase diagram). In the inset a schematic view of
          the samples is shown. Panel c). The anisotropy of the
          AF layer extracted through numerical analysis of the
          phase diagram. The line is a fit to the data based using
          Eq.~1 revealing a finite size effect of the AF
          anisotropy.}
          \label{Fig1}
          \end{figure}

The  samples have been investigated at room temperature by using
the magneto-optical Kerr effect in a longitudinal geometry. In
Fig.~\ref{Fig1}a the azimuthal dependence of EB field(H$_{EB}$)
and coercive field (H$_c$) are shown for a representative sample.
This provides the orientation of the uniaxial and unidirectional
anisotropy induced during growth in an applied magnetic field,
defined as $\phi$=0. The hysteresis loop at this orientation
further provides the coercive and EB fields for each sample. Both
quantities are plotted in Fig.~\ref{Fig1}b as a function of the AF
layer thickness (t$_{AF}$). Fig.~\ref{Fig1}b presents the
so-called {\textit{phase diagram}}  for the EB and coercivity
which is of crucial importance for defining the basic microscopic
mechanism for EB.

The characteristic behavior of the NiFe/IrMn system is very
similar to the
 prediction of the M\&B model~\cite{meiklejohn:1962,rz:2008}.
Experimentally, two regions are clearly visible in the phase
diagram: a) a first region from {t$_{AF}$=0~\AA} to
{t$_{AF}$=25~\AA}where the EB field vanishes and coercive field is
strongly enhanced and b) a second region with t$_{AF}$ higher than
25~\AA~where the EB occurs and the coercive field is reduced.
Within the M\&B model the AF is supposed to rotate under the
torque created by the interfacial exchange coupling, therefore,
transferring  anisotropy energy into the F layer. This is seen as
a peak feature of the H$_c$ with a sharp upturn at {t$_{AF}
\approx 10~\AA$}. Above the critical thickness for EB, the AF is
rigid as a whole, acquiring slight deviation from its equilibrium
position during the magnetization reversal. Experimentally, this
is revealed by an abrupt onset of the H$_{EB}$  which is
accompanied by a decrease of the H$_c$. Interestingly, this
particular behavior of H$_c$ and H$_{EB}$ across the critical AF
thickness is not predicted by the Mauri
model~\cite{maurimodel:1987}. Therefore, the experimental data
(Fig.~\ref{Fig1}a,b) clearly confirm the validity of the M\&B
mechanism of EB for these
bilayers~\cite{rz:2008}.  

Nevertheless, deviations from the M\&B model are still important.
Above the AF critical thickness the coercive field is still
enhanced. For instance, far from the critical thickness, at
{t$_{AF}$=35~\AA} the coercive field is about 50~Oe. This value is
much higher as compared to the coercive field of
permalloy(Py$\equiv$Ni$_{81}$Fe$_{19}$) layer, which for our
samples is about 3~Oe (the experimental point at t$_{AF}$=0, in
Fig.~\ref{Fig1}b). In order to account for this enhanced
coercivity, a new model Spin Glass (SG) model was recently
introduced which suggests that a magnetically disordered interface
may promote enhanced coercivity at the expense of EB field. The
main assumption of the SG model is that the AF anisotropy is
reduced at the interface allowing the formation of frozen and
rotating  AF spins which further affects the coupling strength and
mediate coercivity into the F layer~\cite{radu:jpcm:2006}.

We use this model to analyze the experimental data. First we
simulate the azimuthal dependence (Fig.~\ref{Fig1}a) of the
H$_{c}$ and H$_{EB}$ for an AF thickness of t$_{AF}$=35~\AA. 
The sample was rotated around its normal in 5~degree steps. For
each orientation a hysteresis loop was measured which further
provided the H$_c$ and  H$_{EB}$. The unidirectional behavior for
the EB field is clearly seen as a major $\sin{\phi}$ behavior(down
triangles in Fig.~\ref{Fig1}a). Along the applied field direction
during growth the coercive field is enhanced at $\phi=0$ and
$\phi$=180 deg. The parameters extracted form the
simulations(lines in Fig.~\ref{Fig1}a) are J$_{EB}$=0.185 J/m$^2$,
f=90~\%, and $\gamma$=7 deg, where, J$_{EB}$ is the interfacial
coupling constant, f is a conversion factor and $\gamma$ is the
disordered layer anisotropy orientation~\cite{radu:jpcm:2006}. The
origin of this disorder layer is related to the symmetry breaking
at the interface as well as to the chemical roughness and
inter-diffusion~\cite{radu:jpcm:2006,mccord:2008,camarero:2009}.
Using the f and J$_{EB}$ values extracted above we have further
simulated the phase diagram shown in Fig.~\ref{Fig1}b by reducing
the AF anisotropy to match the measured hysteresis loops for
different AF thicknesses. This  provides the averaged AF
anisotropy constant as a function of AF thickness which is
depicted in Fig.~\ref{Fig1}c). Note that the absolute values of
the AF anisotropy is made possible due to the design of the
samples, where only the AF thickness is varied, keeping the
interface and
 the magnetic properties of the F layer unaffected  by varying only the AF thickness.

\begin{figure}[ht]
          \includegraphics[clip=true,keepaspectratio=true,width=1\linewidth]{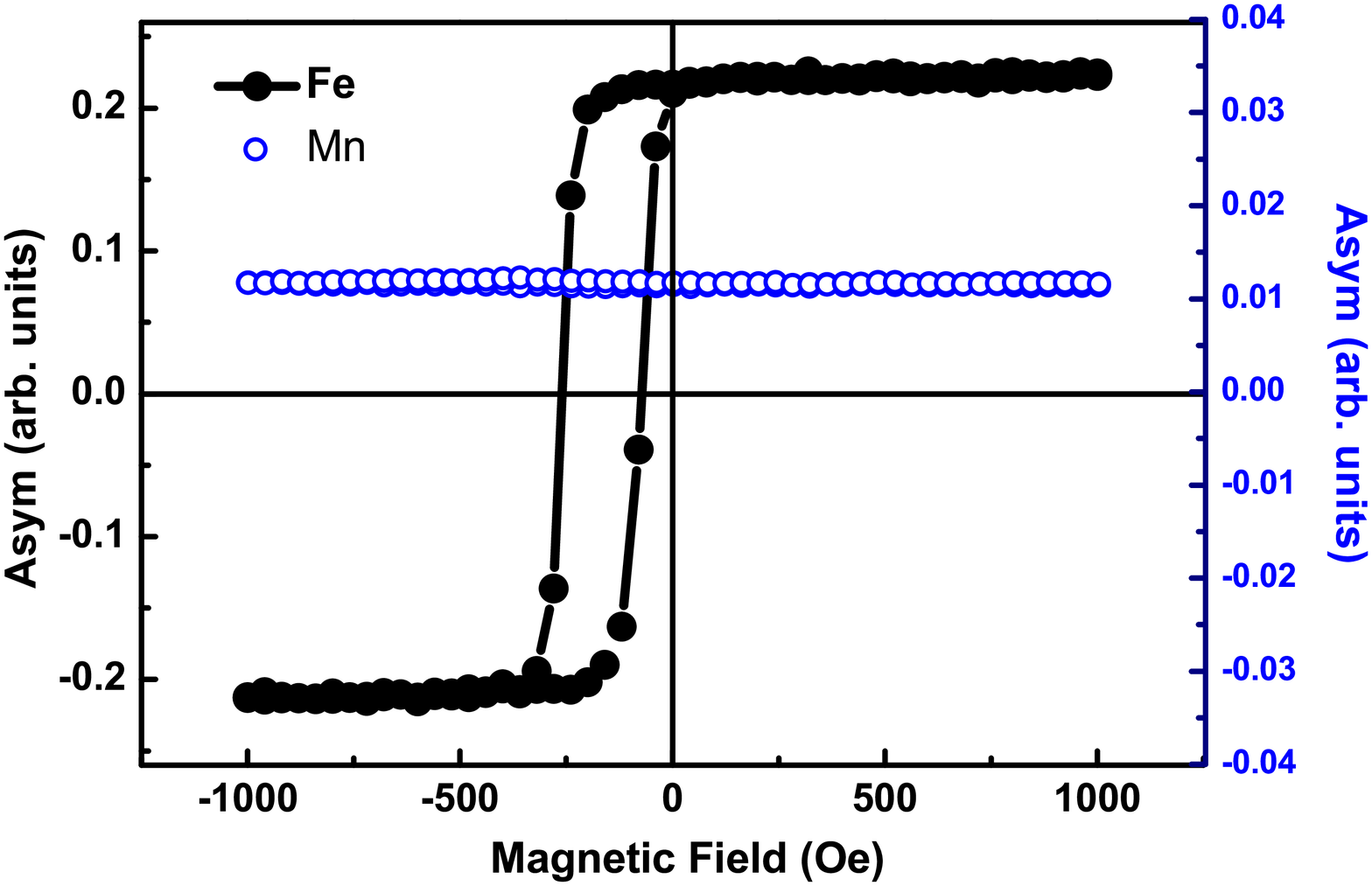}
          \caption {
           Vertically resolved element specific hysteresis  loops
          measured at the L$_3$ resonant energy of Fe (filled circles) and Mn (open circles)
           for the sample measured by soft x-rays (see Fig.~\ref{Fig3}).} 
          \label{Fig2}
          \end{figure}

Strikingly, the AF anisotropy exhibits a finite size effect. When
the dimensions of the magnetic materials are reduced towards the
critical correlation lengths for which  long range order cannot be
sustained, the AF ordering temperature is reduced with respect to
the bulk value~\cite{Jensen:2006}. The ordering temperature can be
related  to the geometric confinement of the magnetic energy, via
scaling laws~\cite{Jensen:2006,ambrose:1996,weschke:2004,he:2007}.
Here we propose  a similar power law (see
Ref.~\cite{ambrose:1996}) for the anisotropy constant:
\begin{equation}
\frac{K_{AF}^\infty-K_{AF}(t_{AF})}{K_{AF}^\infty}=\left(\frac{\xi}{t_{AF}}\right)^\lambda,
\end{equation}
where, $K_{AF}^\infty$ is the bulk AF anisotropy, $\xi$ is the
correlation length at the measuring temperature, and $\lambda$ is
the so called shift exponent for finite-size scaling. This seems
justified since the critical temperature is characterized by the
disappearance of the AF magnetic anisotropy.  Fitting the
experimental AF anisotropy data (see Fig.~\ref{Fig1}e) leads to
the following parameters: {$\xi=10\pm~0.13$~\AA}, $\lambda=2.14
\pm 0.28$, and $K_{AF}^\infty=(1.13~\pm~0.5)~\times~10^{5}~J/m^3$.
The $\xi$ parameter corresponds to the critical thickness for the
onset of the AF anisotropy which further suggests that the N\'eel
temperature of a of {10~\AA} thick IrMn film is 300~K. This is
also clearly seen in the Fig.~\ref{Fig1}b as a sharp upturn of the
coercive field. The value of the shift exponent $\lambda$   is
related to the critical exponent of the correlation length as
$\lambda = 1/\nu$. According to Jensen and
Bennemann~\cite{Jensen:2006}, the non-universal parameter
$\lambda$ does not agree with $1/\nu$ but is related to the
coupling constants in the thin  films. Most importantly, this
analysis provides the average anisotropy constant for thicker
films. In the past  only the possibility  to extract the AF
anisotropy at a critical AF
thickness~\cite{mauri:1987,steenbeck:2004} for EB was explored. In
the light of finite-size effects, the AF anisotropy extracted at a
reduced thickness as in Ref~\cite{mauri:1987} underestimates its
the absolute value.

In order to provide further confidence for the EB mechanism
assumed above, we concentrate now on the dual behavior of the AF
interfacial components accounted for by the conversion factor,
\emph{f}. An unity value for the conversion factor indicates an
ideal interface with no rotating UC AF spins , whereas a zero
value for \emph{f}~\cite{radu:jpcm:2006}.
would translate in a large fraction of rotating AF spins and
vanishing number of frozen-in AF UC spins.


\begin{figure}[ht]
          \includegraphics[clip=true,keepaspectratio=true,width=1\linewidth]{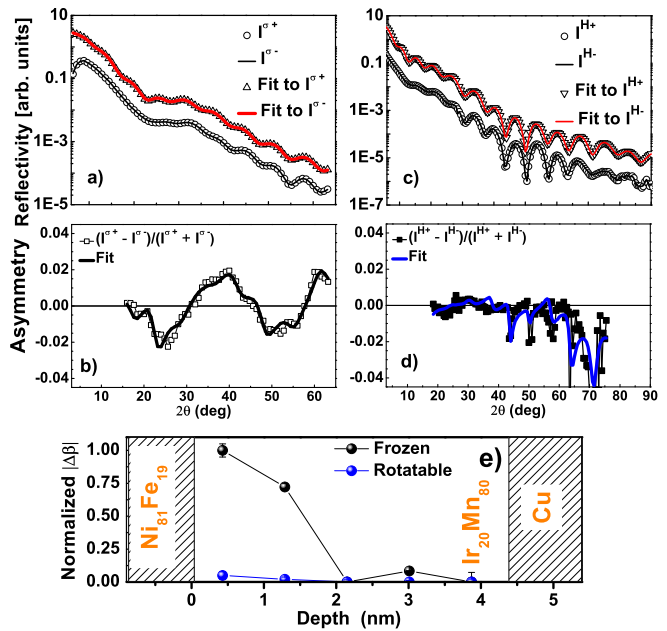}
          \caption{Panel a). The element specific reflectivities
          measured in a saturation field with positive and
          negative helicities at the L$_3$ resonant energy of Mn~(639
          eV). Panel b). The asymmetry extracted from the reflectivities shown in
          pannel a). Panel c). The element specific reflectivities
          measured  with positive  helicity at the L$_3$ resonant energy
          of Mn and for two opposite magnetic fields.
          Panel d). The asymmetry extracted from the reflectivities shown in
          pannel c). Fits to the  experimental data are shown in
          all pannels.
          Panel e). Depth dependence of the frozen and rotating AF (Mn)
          uncompensated spins.}
   \label{Fig3}
 \end{figure}

To probe the UC AF components we have measured element specific
hysteresis loops (Fig.~\ref{Fig2}) and reflectivities
(Fig.~\ref{Fig3}a,c) for a representative sample.
 XRMR measurements
were performed at the UE46 High Field End Station
(Fig.~\ref{Fig3}a,b) and the ALICE
diffractometer~\cite{grabis:2003} at beamline UE56-2
(Fig.~\ref{Fig3}c,d). 
 By tuning the energy  of the incident radiation close to the Mn L$_3$ absorption edge,   we
have measured reflectivity curves which further allow us to select
the scattering conditions for a maximum magnetic contrast at
constant angle of incidence~\cite{radu:jmmm:2006}. This is
achieved by analyzing the asymmetry curves which are of two types:
asymmetry at constant field measured for positive and negative
helicities, and at constant helicity measured for positive and
negative external fields:
$(A^H,A^\sigma)=(I^{(H,\sigma)}-I^{(H,\sigma)})/(I^{(H,\sigma)}+I^{(H,\sigma)})$.
For instance,   Fig.~\ref{Fig3}b,d   shows   the asymmetry at
constant saturation field (A$^H$). We observe an oscillating
magnetic asymmetry as a function of incident angle. In order to
record a hysteresis loop we set the detector angle 2$\theta$ to 35
degrees which provides significant magnetic contrast and
sufficient reflected intensity.
Vertically resolved element
specific hysteresis loops (VR-ESHL) are plotted in Fig.~\ref{Fig2}
for both Fe and Mn resonant energies. We observe that the
ferromagnetic layer behaves as expected. It shows a horizontal
shift of the hysteresis loop and is symmetrically centered with
respect to the magnetization axis. The VR-ESHL at the Mn resonant
energy is very different. Practically, it is completed displaced
with respect to the magnetization axis with a very weak vertical
opening. This suggests that the UC AF component is mainly frozen.
The rotating fraction of the AF UC component is barely visible in
this curve.

To determine the depth profile and the relative fraction of the
frozen and rotating UC spins we have  analyzed the reflectivities
and asymmetry curves for both frozen (Fig.~\ref{Fig3}a,b) and
rotating (Fig.~\ref{Fig3}c,d) UC spins. This is achieved by
fitting the experimental data with an algorithm based on the Zak's
formalism~\cite{zak:1990,valencia:2008}. The results are shown in
Fig.~\ref{Fig3}e. There the relative variation of the magnetic
absorption coefficients for frozen and rotating UC spins are
plotted as a function of depth. We observe that the frozen UC
component contributes dominantly to the UC spins and that it
extend deeper into the AF layer~\cite{rz:2008,bruck:2008}.  The
rotating UC spins are contributing about ten time less to the UC
spins, as compared to the frozen UC spins. They also appear to be
located closer to the interface. Interestingly, the  fraction of
frozen ) UC spins ($F/(R+F)$ is about 90~\% which correlates well
with the f-factor assumed by the SG Model.

In conclusion, we have studied the thickness dependence of the
exchange bias and coercive fields for a nearly ideally behaved
NiFe/IrMn system. Through numerical analysis of the hysteresis
loops within the SG model we have deduced the absolute value of
the AF anisotropy constant. We have observed that it exhibits a
finite-size effect as a function of AF thickness. This provides an
unprecedented opportunity to study anisotropy constants of AF thin
films. To date this cannot be achieved by any other means.
Utilizing XRMR technique we have directly probed the uncompensated
interfacial spin components. The frozen UC spins extend deeper
into the AF film, whereas the rotating ones are located closer to
the interface. This supports a microscopic mechanism for exchange
bias based on interfacial spin disorder.


We gratefully acknowledge Dr. Willy Mahler for providing excellent
technical support during the measurement at UE 56 (BESSY). The
ALICE diffractometer is funded
through the BMBF Contract No. 05KS7PC1.\\
\\
$\ast$florin.radu@helmholtz-berlin.de


\end{document}